\documentclass[preprint]{revtex4}
\usepackage{epsfig,latexsym,amssymb}
\usepackage{latexsym}
\usepackage{amssymb}
\usepackage{amsfonts}
\usepackage{comment}
\usepackage{graphicx}
\usepackage{verbatim}

\newcommand{\be}{\begin{equation}}
\newcommand{\ee}{\end{equation}}
\newcommand{\bea}{\begin{eqnarray}}
\newcommand{\eea}{\end{eqnarray}}

\begin{document}

\title{Cosmic Microwave Background Dipole Asymmetry could be explained by Axion Monodromy Cosmic Strings}

\author{Qiaoli Yang$^{1}$\footnote{qiaoli\_yang@hotmail.com}, Hongbiao Yu$^{1}$, and Haoran Di$^{2}$}
\affiliation{$^1$Physics Department and Siyuan Laboratory, Jinan University, Guangzhou 510632, China\\$^2$School of Physics, Huazhong University of Science and Technology, Wuhan 430074, China}

\begin{abstract}
Observations by the Wilkinson Microwave Anisotropy Probe and the Planck mission suggest a hemispherical power amplitude asymmetry in the cosmic microwave background, with a correlation length on the order of the size of the observable Universe.
We find that this anomaly can be naturally explained by an axion-like particle (ALP) cosmic string formed near our visible Universe.
The field variation associated to this cosmic string creates particle density fluctuations after inflation, which consequently decay into radiation before the Big Bang Nucleosynthesis (BBN) era and resulted in the observed power asymmetry.
We find in this scenario that the hemispherical power amplitude asymmetry is strongly scale dependent: $A(k)\propto {\rm exp}(-kl)/k$.
Admittedly, typical inflation models predict a relic number density of topological defects of order one per observable Universe and so in our model the cosmic string must be tuned to have an impact factor of order $1/H_0$.
Interestingly, the constraints based on purely cosmological considerations also give rise to a Peccei-Quinn scale $F_a$ of order $10^3$ larger then the Hubble scale of inflation $H_I$.
Assuming $H_I\sim 10^{13}$GeV, we then have an ALP with $F_a\sim 10^{16}$GeV, which coincides with the presumed scale of grand unification.
As we require ALP decays occur before the BBN era, which implies a relatively heavy mass or strong self-coupling, and considering that the associated potential should break the shift symmetry softly in order to protect the system from radiative corrections, we also conclude that the required ALP potential should be monodromic in nature.
\end{abstract}

\maketitle


\section{Introduction}
The predictions of inflationary theory \cite{Guth:1980zm,Albrecht:1982wi,Linde:1981mu} are remarkably consistent with cosmological observations \cite{Komatsu:2010fb,Ade:2013zuv}.
In the typical form of the theory, the Universe experienced a period of exponential expansion driven by a scalar field with a flat potential; the inflaton, and therefore, the Universe we observe today was once in a region with an established causal contact.
Spatial curvature and the density of exotic relics were diminished by this exponential expansion, and the quantum fluctuations of the inflaton field provided the seeds of the large-scale cosmic structures we observe today.
Furthermore, the corresponding primordial density fluctuations can be observed by measuring the cosmic microwave background (CMB) anisotropy.

However, despite these successes the Wilkinson Microwave Anisotropy Probe (WMAP) and the Planck mission have also found that large-scale anomalies exist with respect to ordinary inflationary theory, notably including a hemispherical power asymmetry with correlation length on the order of the size of the current Universe \cite{Eriksen:2003db,Hansen:2004vq,Land:2005ad,Eriksen:2007pc,Hansen:2008ym,Hoftuft:2009rq,Paci:2013gs,Cai:2013gma,Cai:2015xba,Akrami:2014eta,Ade:2015hxq,Ade:2013nlj,Donoghue:2004gu,Hanson:2009gu,Bennett:2010jb,Hirata:2009ar,Flender:2013jja,Adhikari:2014mua,Aiola:2015rqa,Yang:2016wlz}.
 This power asymmetry also has a strong scale-dependence which further defies explanation.

To understand these observations we propose a model with a pseudo-goldstone field $a$, in addition to the inflaton, with an associated spontaneously broken U(1) symmetry, which played an essential role in the formation of the CMB anisotropy.
If the spontaneous breaking of this U(1) symmetry created string-like topological defects, one of which was near our observable Universe, then this topological defect would result in a primordial density variation of bosonic particles around it.
This density variation would then naturally imbue the CMB power dipole asymmetry with a scale-dependent spectrum.
The particular dynamic properties of the pseudo-goldstone field we require occur specifically in models of axion-like particles (ALPs) with monodromic potentials.

Admittedly, typical inflation models predict a relic number density of topological defects of order one per observable Universe and consequently the position of the defect needs to be tuned to lie near our observable Universe.
So the cosmic string impact factor $l\sim 1/H_0$ (see FIG. 1) in this model.
In addition, this model will only focus on resolving the CMB-dipole anomaly in which we are most interested, mainly because it could indicate the presence of pre-inflationary physics, and therefore other CMB spectrum anomalies will not be considered here.

Axions \cite{Weinberg:1977ma,Wilczek:1977pj,Vysotsky:1978dc,Kim:1979if,Shifman:1979if,Zhitnitsky:1980tq,Dine:1981rt,Sikivie:1982qv,Berezhiani:1990sy,Berezhiani:1992rk} were originally proposed to solve the strong CP problem in quantum chromodynamics (QCD), via a new Peccei-Quinn symmetry \cite{Peccei:1977hh} which is spontaneously broken and gives rise to a Goldstone particle, the ``axion".
QCD instanton effects create an explicit U(1) symmetry-breaking potential, which then gives a nonzero mass to the axion.
It was later discovered that in string theory, ALPs \cite{Davis:2005jf,Dolan:2017vmn,Svrcek:2006yi,Douglas:2006es,witten:1986,Becker:1995kb} can arise from the compactifications of gauge field over the cycles of the compactified extra dimensions.
Owing to the complexities of the compactified manifold, there can be many different ALP species in the effective four-dimensional theory.
Notably, ALPs in string theory can also form cosmological strings \cite{Davis:2005jf,Dolan:2017vmn}.

Many models predict a decay constant $F_a$ on the order of the grand unified theory (GUT) scale of $10^{16}$ GeV \cite{Davis:2005jf,Svrcek:2006yi}.
In addition, string instantons such as worldsheet or brane instantons can create a potential with a discrete shift symmetry similar to that of QCD axions.
As this potential depends exponentially on the instanton action, and given the uncertainties associated to string instantons, the possible mass range of string axions or ALPs is very large.

Of particular interest are axions with monodromic potentials, which spontaneously break shift symmetry and have been proposed to serve as the inflaton \cite{Silverstein:2008sg,Kaloper:2008fb,McAllister:2008hb,Flauger:2009ab,Conlon:2011qp,Hebecker:2014eua,Baumann:2014nda,Palti:2014kza,Marchesano:2014mla,Escobar:2015ckf,Hebecker:2015tzo}.
In this scenario, the monodromy potential is non-periodic and multi-branched, so an ALP field with sub-Planckian symmetry-breaking scale can drive inflation via a trans-Planckian field excursion.
Interestingly, although the ALP we consider in the current context is not the field directly driving inflation, the potential required due to compatibility with observations nevertheless seems to be monodromic in nature.

\section{Axion-like particles and pre-inflationary topological defects}
We now consider a general ALP with a discrete shift symmetry $a\to a+2\pi F_a$ preserving potential $V(a)$ and a monodromy potential $U(a)$.
The $V(a)$ potential can be generically written as
\bea
V(a)= \Lambda^4[1- {\rm cos}({a\over F_a})]~~.
\eea
The $V$ term resulting mass $m_0$ and quartic self-coupling $\lambda_0$ are
\be
m_0={\Lambda^2\over F_a},~~\lambda_0=-{\Lambda^4\over F_a^4}~~,
\ee
respectively, where $\Lambda$ is generated by the nonperturbative effects of string instantons and as such is mode-dependent \cite{ Svrcek:2006yi}.
One estimation is $\Lambda^2\sim M_{Pl}\Lambda_0 e^{-S/2}$, where $S$ is the instanton action and $\Lambda_0$ typically varies over a range: [$10^{4}{\rm GeV},~~10^{18}{\rm GeV}]$.
The instanton action $S$ is model-dependent, with a typical value in the several-hundred range \cite{Visinelli:2018utg}.
Consequently, if $F_a$ is order of the GUT scale, the axion mass is typically sub-GeV.

\begin{figure*}
\begin{center}
\includegraphics[width=1.0\textwidth]{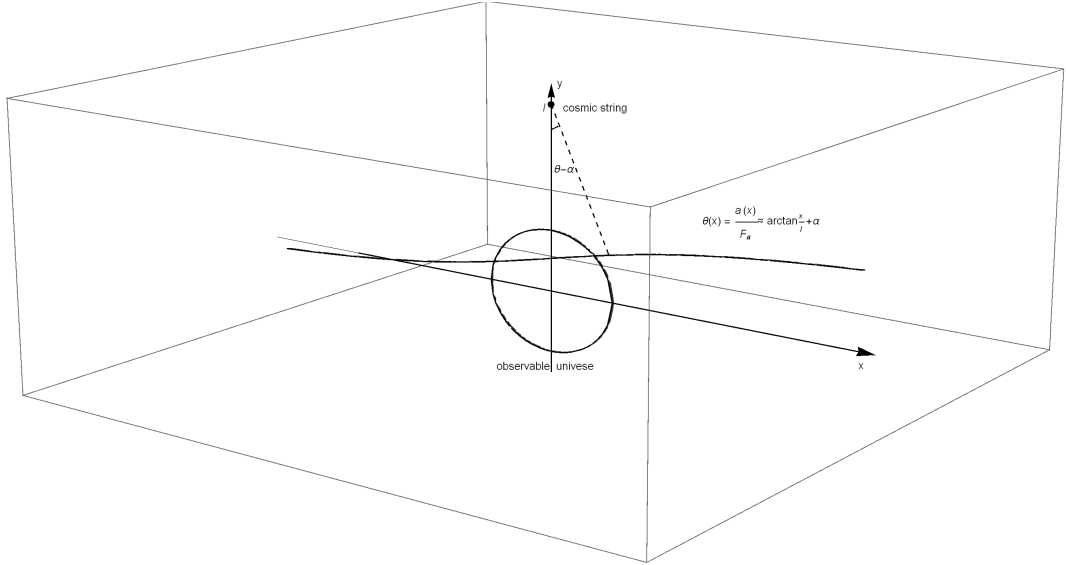}
\caption{A schematic illustration of the amplitude fluctuation induced by a cosmic string.
$l$ denotes the relative position between the nearest cosmic string and our Universe.
The ALP field varies around the cosmic string, and consequently after Hubble friction is sufficiently decreased, a boson condensate is produced.
$\alpha F_a$ is the condensate background and $F_a $arctan[x/l] is the fluctuation of the condensate.}
\end{center}
\end{figure*}

The $U$ term appears in the axion monodromy scenario, which is proposed to extend the effective axion field range where the potential reaches a new configuration after being traversed by a closed loop in the axion configuration space.
The shift symmetry is only spontaneously broken by the $U$ term so the radiative corrections are still under control \cite{Silverstein:2008sg,Kaloper:2008fb,McAllister:2008hb,Dolan:2017vmn}.
Monodromies are common phenomena in string compactifications, but the explicit construction of axion monodromy models is delicate.
The monodromy potential is often parameterized using an exponent $p$ \cite{Baumann:2014nda}, via
\be
U(a)=\mu^{4-p}a^p~,
\ee
where typically $p\in [0.2,~~5]$ for inflation.
In this paper we are interested in $p=2$ case, which typically results from axion coupling with a four-form or on a pair of seven-brane and gives $\mu\sim 10^{13}$GeV \cite{Kaloper:2008fb}, and also consider a hypothetical $p=4$ case.

Immediately after the temperature of the Universe dropped below the ALP symmetry breaking scale, the Hubble rate was high and therefore the potential term $V'(a)+U'(a)$ was sub-dominant, so that equation of motion of the ALP was
\be
(\partial_t^2+{3H}\partial_t+{k^2\over R^2})a(\vec k ,t)=0~~,
\label{horizon}
\ee
where $R$ is the scale factor, and $H$ is the Hubble factor.
For a comoving wavelength $2\pi(k/R)^{-1}\gg H^{-1}$, the solution is $a(\vec k, t)\sim a(\vec k)$, which is frozen out by causality.
For a comoving wave length $2\pi(k/R)^{-1}\ll H^{-1}$, the modes have a decreasing amplitude owing to the Hubble term; therefore, such subhorizon modes are typically diluted by the expansion of the Universe.

If however a topological defect appears, the field fluctuations around the defect are protected by topology as long as the defect is inside the horizon.
Consider (see Fig.1) a cosmic string appearing near our observable Universe; the variation of the ALP field around a cosmic string is $\delta a=2\pi F_a$, and so the initial field configuration is then
\bea
\Delta \theta=\theta-\alpha={a(x)\over F_a}-\alpha=\arctan{x\over l-y}~~,
\eea
where $l$ is the distance from the defect to the observable Universe.
For the region of interest, the ALP field is configured as
\bea
a(\vec x)\approx \arctan ({x\over l})F_a+\alpha F_a= \bar a_0 \arctan(\vec k\cdot \vec x)+ \Sigma~,
\eea
where $\vec k=( 1/l,0,0)$ defines the preferred direction, $\bar a_0=F_a$ is the amplitude of the fluctuation, and $\Sigma=\alpha F_a$.
As the field fluctuation can be expanded as
\bea
\arctan(kx)\propto\int (e^{-|k'/k|}/k')e^{-ik'x}dk'~~,
\label{scale}
\eea
the fluctuation is dominated by long-wavelength modes with a pivot scale: $(1/k')\gtrsim (1/k)=l$.
During inflation, $\Delta \theta$ remains constant if the topological defect is within the horizon.
If the defect crosses the horizon, fluctuation modes with a superhorizon wavelength are frozen by causality.

Before the axion field started to oscillate, the axion potential terms were sub-dominant so the axion field was protected by the shift-symmetry.
After inflation, the Universe was reheated by the inflaton and later entered the radiation-dominated era.
When the Hubble rate decreased to a scale at which the potential terms become comparable to the Hubble friction, the ALP field started to oscillate and became a condensate of nonrelativistic cold bosons.
The newly created boson condensate had a primordial fluctuation $\Delta a=\Delta \theta F_a$ across the observable Universe.
The cosmic string itself and a newly formed domain wall at $\theta=0$ are well outside the observable region, as $\Delta \theta<\alpha$.

When $p=4$, the created bosons will have an effective mass in this condensed mode due to their self-coupling.
The energy density decreases as $R^{-4}$ instead of $R^{-3}$ because the effective mass also decreases during the dilution of the condensate.
The energy density created by the inflaton decreases as $R^{-4}$ during the radiation dominated era and thus, the energy density ratio between the two components is constant.

When $p=2$ the ALP energy density decreases as $R^{-3}$ and the radiation density created by the inflaton decreases as $R^{-4}$, and so the energy density ratio between the two component scales as $\rho_a/\rho \propto R$ before the particles decay into radiation.

\section{Decay of the axion-like particle}

{\itshape $p=4$}:
Owing to the self-interaction term $\lambda$, the effective mass in an ALP condensate is
\be
m_{ef}^2\approx\lambda {\bar a}^2~~,
\label{efm}
\ee
where $\bar a $ is the averaged field strength.
The decay rate in the condensate is therefore
\be
\Gamma={g\over 4\pi}{m_{ef}^3\over F_a^2}~~,
\ee
where $g$ is a model-dependent factor.
If $\lambda \gtrsim  4\pi/g$, the bosons will decay as soon as they begin to oscillate.

The momentum of the newly created bosons is negligible so the energy density of the condensate is
\bea
\rho_{a}\approx {1\over 2}m_{ef}^2{\bar a}^2={1\over 2}\lambda{\bar a }^4~~,
\label{sigma}
\eea
which decreases in proportion to $1/R^4$ despite being non-relativistic.
The total energy density at a time $t\sim 1/\Gamma$ is dominated by the inflaton-sourced radiation:
\bea
\rho={g_*\pi^2T_B^4\over30}[{R(t_B)\over R(t)}]^4~~,
\label{tot}
\eea
where we use the BBN era as a reference time so $t_B$ is the time of the Big Bang Nucleosynthesis era, $T_B$ is order of MeV, and $g_*$ is the number of effective degrees of freedom during that era.
The energy density ratio between the two components is a constant in the radiation dominated era:
\be
{\rho_a\over\rho}\sim {15\bar a_0 ^2\over g_*\pi^2T_B^4t_B^2},
\label{ratio1}
\ee
where $a_0$ is the initial field strength.
If $F_a$ is of order the GUT scale, the energy density ratio is of order $10^{-3}$.
Notably the mass of this ALP in vacuum is small; typically sub-GeV, which makes laboratory direct detection possible.

{\itshape $p=2$}:
As the decay rate of the ALP is
$
\Gamma={(g /4\pi)}{(m^3/ F_a^2)}
$,
where we will take $g\sim {\cal O}(1)$ in the following discussions, we have
\be
\Gamma>H_B=\sqrt{{8\pi\over 3G} {g_*\pi^2T_B^4\over30}}~~,
\ee
In addition, the energy density of the ALP at a time $t$ just before their decay is:
\bea
\rho_{a}\approx m^2F_a^2[{R(t_1)\over R(t)}]^3~~,
\label{sigma}
\eea
where $t_1\approx 1/m$ is the time that the ALP field started to oscillate and $t\approx 1/\Gamma$ is the lifetime of the ALP.
The energy density ratio is
\be
{\rho_{a}\over\rho}\sim1.7\times10^{-37}{\rm GeV}^{-2}{F_a^3\over  m}~,
\label{ratio2}
\ee
when the ALP decays.
If $F_a$ is on the GUT scale, $m\sim \mu\sim 10^{13}$GeV, and the ratio is order of $10^{-2}$.

In either case, the ALPs decayed into ordinary radiation before the BBN era therefore they will not contribute to the exotic energy density or modify the photon-baryon ratio etc.
\section{The CMB anisotropy spectrum}
\begin{figure}
\begin{center}
\includegraphics[width=0.9\textwidth]{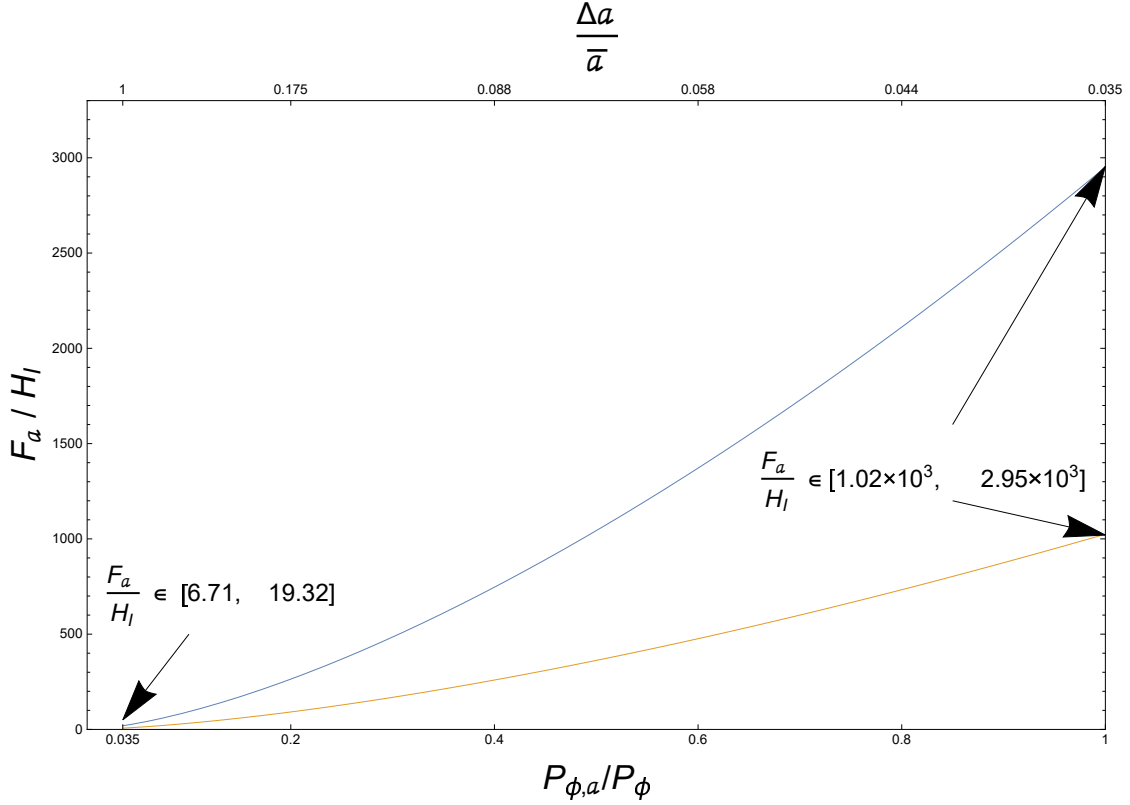}
\caption{The parameter space of the proposed scenario, where the area between the blue and yellow lines is allowed.}
\end{center}
\end{figure}
The observed dipole power asymmetry is
\be
\Delta T(\hat n)=(1+A\hat p\cdot\hat n)\Delta T_{iso}(\hat n)~~,
\ee
where $\hat n$ is a unit vector pointing in the observation direction, $\hat p$ is the dipole asymmetry preferred direction, and $A\approx0.072$ on large scales (from the WMAP and the Planck \cite{Donoghue:2004gu,Hanson:2009gu,Bennett:2010jb}).
On small scales ($\ell=601-2048$), however, $A<0.0045$ \cite{Hirata:2009ar,Flender:2013jja,Adhikari:2014mua,Aiola:2015rqa}; thus, the dipole asymmetry has a very strong scale dependence.

When an initial fluctuation of the ALP field is created across the observable Universe, the quantum nature of the field gives rise to an additional amplitude fluctuation: $\delta a(k)=(H_I/2\pi)|_{(k/R)=H_I}$.
As the ALP field is effectively massless during inflation, the spectrum of this fluctuation is flat.

After inflation and when the Hubble rate has decreased to the order of the effective mass, the ALP field will start to oscillate; the energy density of the ALP field is written as $\rho(\vec x)=({1/2})m^2\bar a^2(\vec x)$, where the bar indicates the amplitude of the oscillation.
Because this fluctuation is small in comparison to the amplitude itself, the energy density contrast is
\bea
{\delta\rho_{a}(\vec x)\over \rho_{a}(\vec x)}\approx2{\delta a(\vec x)\over\bar a(\vec x)}\sim {H_I\over \pi\bar a(\vec x)}~~.
\label{contrast}
\eea
where we have used the condition that the fluctuation spectrum is flat.
The curvature perturbation is then \cite{Lyth:2001nq}
\be
P_{\Phi,a}\approx ({\rho_{a}\over\rho})^2({H_I\over\pi\bar a})^2~.
\label{spectrum}
\ee
On the other hand, assuming $\rho_a/\rho\ll 1$, we have $\delta P_{\Phi,a}\propto 2\delta a\bar a$.
The power spectrum asymmetry on large scales is therefore
\bea
{\Delta P_{\Phi}\over P_{\Phi}}={2\Delta a\over\bar a}{P_{\Phi,a}\over P_{\Phi}}\approx0.07~~.
\label{asymmetry}
\eea
Considering the scale dependence, we have
\bea
{\Delta a\over \bar a}={\bar a_0 \over \Sigma}\arctan(\vec k\cdot \vec x_{dec}),
\label{asys}
\eea
where $x_{dec}$ is the decoupling scale in comoving coordinates.
The RHS of Eq.(\ref{asys}) can be transformed into a harmonic mode basis (see Eq.(\ref{scale})), so that we have
\be
A(k)\propto {e^{-kl}\over k}~~.
\ee

Observations \cite{Donoghue:2004gu,Hanson:2009gu,Bennett:2010jb} have found $A\sim 0.07$ and $\delta A\sim0.02$ at large scales $({\ell}<64)$, however at small scales ($\ell>600$), the Sloan Digital SKy Survey (SDDS) quasar sample observations and the CMB high-$\ell$ spectrum measurement found the power asymmetry is $|A|<0.0045$ for $\ell >600$ \cite{Hirata:2009ar,Flender:2013jja,Adhikari:2014mua,Aiola:2015rqa}.
Consequently it has been suggested that the amplitude of the power asymmetry could be strongly scale-dependent.
Generally speaking, the asymmetry seems large on large scales and suppressed on small scales.
At present there are a number of differing proposals with different scale-dependent power spectra \cite{Erickcek:2008sm,Watanabe:2010fh,Dai:2013kfa,Lyth:2013vha,Liu:2013kea,Mazumdar:2013yta,Cai:2013gma,Lyth:2014mga,Cai:2015xba,Jazayeri:2017szw}, and so future observations such as the Omniscope \cite{Li:2019bsg} could be very important for this issue.

The inhomogeneity of the ALP field also creates perturbations in the CMB power spectrum through the Grishchuk-Zel'dovich effect \cite{L.Grishchuk}.
The temperature fluctuation is \cite{Erickcek:2008jp}:
\be
{\Delta T\over T}(\hat n)=0.066\mu^2{\rho_{a}\over\rho}({\bar a_0\over \Sigma})^2(\vec k\cdot \vec x_{de})^2+..~~,
\ee
where $\mu=\hat k\cdot \hat n$, and $\hat n$ is a unit vector pointing along the line of sight.
As the leading term is quadratic, the constraint is from the CMB quadrupole spherical-harmonic coefficient.
The resulting bound is \cite{Yang:2016wlz}:
\bea
({\rho_{a}\over \rho})({\bar a_0\over \Sigma})^2(\vec k\cdot \vec x_{de})^2\lesssim 4.40\times10^{-4}~.
\eea
By combining this result with Eq.(\ref{asymmetry}), we have
\bea
{\rho_{a}\over\rho}\lesssim 0.36\left(P_{\Phi,a}\over P_{\Phi}\right)^{2}~~.
\label{a1}
\eea

Another consideration is the non-Gaussian contribution from the ALP field.
The fluctuation $\delta a(\vec x)$ is a local Gaussian random variable (see Eq.(\ref{contrast})) but the energy density $\rho_a=(1/2)m^2\bar a^2$ has a contribution quadratic in $(\delta a)^2$.
It implies a non-Gaussian contribution to the density fluctuation, whilst other more complicated non-Gaussian effects should be negligible.
Non-Gaussianity is parametrized by the parameter $f_{NL}$ via $\Phi=\Phi_{gauss}+f_{NL}\Phi_{gauss}^2$.
As $f_{NL}={(5/ 4)}{(\rho/ \rho_{a})}({P_{\Phi,a}/ P_{\Phi}})^2$ \cite{Verde:1999ij,Malik:2006pm,Ichikawa:2008iq}, and the Planck mission indicates that $f_{NL}\lesssim 0.01\%\times10^5$ \cite{Ade:2015ava}, we have
\be
{1\over 8}\left({P_{\Phi,a}\over P_{\Phi}}\right)^2\lesssim {\rho_{a}\over \rho}~~.
\label{a2}
\ee

Fluctuations of particle species density $\delta (n_i/s)\neq 0$ at a fixed total energy density $\delta \rho=0$ are called isocurvature fluctuations.
Generally speaking, QCD axions generate such fluctuations because they are essentially massless in the early Universe hence do not alter the energy density during that era.
Later after they acquire a mass, the energy is transformed from the QCD sector to the axions so the total energy density is conserved and the axion density is compensated by fluctuations of the Standard Model particles such as photons.
In contrast, the energy density of the ALPs in our discussion is generated after the Hubble rate is smaller than the monodromy potential which is decoupled to the Standard Model sector.
The total energy density $\delta \rho\neq0$, and hence this axion field does not generate isocurvature fluctuations.

Finally, as $P_{\Phi}\sim 1.5\times10^{-9}$ \cite{Ade:2015lrj}, Eq.(\ref{spectrum}) leads to
\be
{\rho_{a}\over \rho}\approx1.22*10^{-4}\sqrt{{P_{\Phi,a}\over P_{\Phi}}}\left({F_a\over H_I}\right)~~.
\label{a3}
\ee

By considering the constraints of Eqs.(\ref{a1}), (\ref{a2}), and the relationship of Eq.(\ref{a3}), we find $F_a/H_I\in[6.71, 2.95\times10^3]$ (see Fig. 2).
In addition as $\Delta a/\bar a$ is typically small, ${\cal O}(10^{-1})\sim {\cal O}(10^{-2})$, $F_a/H_I\sim 10^3$.
Assuming $H_I\sim 10^{13}$GeV, we find that $F_a\sim 10^{16}$GeV, $\mu\sim 10^{13}$GeV ($p=2$) or $\lambda\sim {\cal O}(1)$ ($p=4$).

\section{Summary and Discussion}
The hemispherical power asymmetry anomaly in the CMB seems to contradict the inflationary paradigm, as the corresponding correlation length is on the order of the size of our Universe.
In addition, the apparently scale dependent nature of this anomaly further defies explanation.
If however we consider the possible scenario of an ALP cosmic string formed near the observable region of our Universe, with a tuned impact factor $l\sim 1/H_0$, the resulting field fluctuation around the string can naturally explain the observed asymmetry.
Purely cosmological considerations in this scenario suggest a symmetry breaking scale coincident with the GUT scale.
If the observed CMB dipole anomaly scales exponentially, it may furthermore suggest an axion monodromy cosmic string existing near our visible Universe.

\section*{ACKNOWLEDGMENTS}
Q. Yang thanks Yifu Cai, Yu Gao, Maxim Khlopov, Tianjun Li, Danning Li, Hui Liu, Peng Liu, Nick Houston, Anupam Mazumdar and Fanrong Xu for valuable discussions. This work is partially supported by the Natural Science Foundation of China under Grant Number 11305066 and 11875148. This work is funded in part by the Gordon and Betty Moore Foundation through Grant GBMF6210 to Q. Yang.

\end{document}